\documentclass[fleqn,usenatbib]{mnras}

\usepackage{newtxtext,newtxmath}

\usepackage[T1]{fontenc}

\DeclareRobustCommand{\VAN}[3]{#2}
\let\VANthebibliography\thebibliography
\def\thebibliography{\DeclareRobustCommand{\VAN}[3]{##3}\VANthebibliography}

\usepackage{graphicx}	
\usepackage{amsmath}	

\def\beq{\begin{equation}}
\def\eeq{\end{equation}}
\usepackage{calrsfs}
\DeclareMathAlphabet{\pazocal}{OMS}{zplm}{m}{n}

\title[]{Using the motion of S2 to constrain scalar clouds around Sgr A$^*$}
\author[GRAVITY Collaboration]{GRAVITY Collaboration \thanks{GRAVITY is developed in collaboration by 
MPE, 
LESIA of Paris Observatory / CNRS / Sorbonne Universit\'e / Univ. Paris Diderot 
and IPAG of Universit\'e Grenoble Alpes / CNRS, 
MPIA, 
Univ. of  Cologne,
CENTRA - Centro de Astrof\'{\i}sica e Gravita\c c\~ao, and ESO. Corresponding authors: A.~Foschi 
(arianna.foschi@tecnico.ulisboa.pt), V.~Cardoso (vitor.cardoso@tecnico.ulisboa.pt), \& P.J.V.~Garcia (pgarcia@fe.up.pt)
}:
A.~Foschi$^{1, 2}$,
R.~Abuter$^{3}$,
N.~Aimar$^{4}$, 
P.~Amaro Seoane$^{5, 7, 9, 26}$, 
A.~Amorim$^{1, 10}$, \newauthor
M.~Baub\"ock$^{5, 11}$,
J.P.~Berger$^{12}$,
H.~Bonnet$^{3}$, 
G.~Bourdarot$^{5}$, 
W.~Brandner$^{13}$,
V.~Cardoso$^{1, 6}$,
Y.~Cl\'{e}net$^{4}$, \newauthor
Y.~Dallilar$^{5}$,
R.~Davies$^{5}$,
P.T.~de~Zeeuw$^{14}$,
D.~Defrère$^{24}$,
J.~Dexter$^{15}$, 
A.~Drescher$^{5}$, 
A.~Eckart$^{20, 23}$,\newauthor
F.~Eisenhauer$^{5}$, 
M.C.~Ferreira$^{1}$,
N.M.~F\"orster~Schreiber$^{5}$,
P.J.V.~Garcia$^{1, 2}$, 
F.~Gao$^{5, 16}$,
E.~Gendron$^{5}$, \newauthor
R.~Genzel$^{5,17}$, 
S.~Gillessen$^{5}$, 
T.~Gomes$^{1, 2}$,
M.~Habibi$^{5}$, 
X.~Haubois$^{18}$,
G.~Hei{\ss}el$^{4, 19}$, 
T.~Henning$^{13}$, \newauthor
S.~Hippler$^{13}$, 
S.F.~H\"{o}nig$^{8}$,
M.~Horrobin$^{20}$,  
L.~Jochum$^{18}$,
L.~Jocou$^{12}$, 
A.~Kaufer$^{18}$, 
P.~Kervella$^{4}$,  \newauthor
L.~Kreidberg$^{13}$,
S.~Lacour$^{4}$, 
V.~Lapeyr\`ere$^{4}$, 
J.-B.~Le~Bouquin$^{12}$, 
P.~L\'ena$^{4}$, 
D.~Lutz$^{5}$, 
F.~Millour$^{25}$,
T.~Ott$^{5}$,\newauthor
T.~Paumard$^{4}$, 
K.~Perraut$^{12}$, 
G.~Perrin$^{4}$, 
O.~Pfuhl$^{3, 5}$, 
S.~Rabien$^{5}$, 
D.C.~Ribeiro$^{5}$, 
M.~Sadun Bordoni$^{5}$,  \newauthor
S.~Scheithauer$^{13}$, 
J.~Shangguan$^{5}$,
T.~Shimizu$^{5}$,  
J.~Stadler$^{5, 21}$, 
O.~Straub$^{5, 22}$, 
C.~Straubmeier$^{20}$, 
E.~Sturm$^{5}$, \newauthor
C.~Sykes$^{8}$, 
L.J.~Tacconi$^{5}$, 
F.~Vincent$^{4}$, 
S.~von~Fellenberg$^{5, 23}$,
F.~Widmann$^{5}$, 
E.~Wieprecht$^{5}$, 
E.~Wiezorrek$^{5}$ \newauthor
and J.~Woillez$^{3}$ 
\\
$^{1}$CENTRA - Centro de Astrof\'{\i}sica e
Gravita\c c\~ao, IST, Universidade de Lisboa, 1049-001 Lisboa,
Portugal\\
$^2$Faculdade de Engenharia, Universidade do Porto, rua Dr. Roberto
Frias, 4200-465 Porto, Portugal\\ 
$^3$European Southern Observatory, Karl-Schwarzschild-Stra{\ss}e 2, 85748
Garching, Germany\\
$^4$LESIA, Observatoire de Paris, Universit\'e PSL, CNRS, Sorbonne Universit\'e, Universit\'e de Paris, 5 place Jules Janssen, 92195 Meudon, France\\
$^5$Max Planck Institute for extraterrestrial Physics,
Giessenbachstra{\ss}e~1, 85748 Garching, Germany\\
$^6$Niels Bohr International Academy, Niels Bohr Institute,
Blegdamsvej 17, 2100 Copenhagen, Denmark\\
$^{7}$Universitat Politècnica de València, València, Spain \\
$^{8}$School of Physics \& Astronomy, University of Southampton, Southampton, SO18 4EX, United Kingdom \\
$^{9}$ Kavli Institute for Astronomy and Astrophysics, Beijing, China \\
$^{10}$Universidade de Lisboa - Faculdade de Ci\^encias, Campo Grande,
1749-016 Lisboa, Portugal\\
$^{11}$Department of Physics, University of Illinois, 1110 West Green Street, Urbana, IL 61801, USA\\
$^{12}$Univ. Grenoble Alpes, CNRS, IPAG, 38000 Grenoble, France\\
$^{13}$Max Planck Institute for Astronomy, K\"onigstuhl 17, 
69117 Heidelberg, Germany\\
$^{14}$Leiden University, 2311EZ Leiden, The Netherlands\\
$^{15}$Department of Astrophysical \& Planetary Sciences, JILA, Duane Physics Bldg., 2000 Colorado Ave, University of Colorado, Boulder, CO 80309, USA\\
$^{16}$Hamburger Sternwarte, Universit\"at Hamburg, Gojenbergsweg 112, 21029 Hamburg, Germany\\
$^{17}$Departments of Physics and Astronomy, Le Conte Hall, University
of California, Berkeley, CA 94720, USA\\
$^{18}$European Southern Observatory, Casilla 19001, Santiago 19, Chile\\
$^{19}$Advanced Concepts Team, European Space Agency, TEC-SF, ESTEC, Keplerlaan 1, 2201, AZ Noordwijk, The Netherlands \\
$^{20}$ $1^{\rm st}$ Institute of Physics, University of Cologne,
Z\"ulpicher Stra{\ss}e 77, 50937 Cologne, Germany\\
$^{21}$Max Planck Institute for Astrophysics, Karl-Schwarzschild-Stra{\ss}e 1, D-85748
Garching, Germany\\
$^{22}$ORIGINS Excellence Cluster, Boltzmannstra{\ss}e 2, D-85748 Garching, Germany \\
$^{23}$Max Planck Institute for Radio Astronomy, auf dem H\"ugel 69, D-53121 Bonn, Germany \\
$^{24}$Institute of Astronomy, KU Leuven, Celestijnenlaan 200D, 3001 Leuven, Belgium \\
$^{25}$Université C\^{o}te d'Azur, Observatoire de la  C\^{o}te d'Azur, CNRS, Lagrange, France \\
$^{26}$ Higgs Centre for Theoretical Physics, Edinburgh, UK}
\date{Accepted XXX. Received YYY; in original form ZZZ}

\pubyear{2023}

\begin{document}
\label{firstpage}
\pagerange{\pageref{firstpage}--\pageref{lastpage}}
\maketitle

\begin{abstract}
The motion of S2, one of the stars closest to the Galactic Centre, has been measured accurately and used to study the compact object at the centre of the Milky Way. It is commonly accepted that this object is a supermassive black hole but the nature of its environment is open to discussion. Here, we investigate the possibility that dark matter in the form of an ultralight scalar field ``cloud'' clusters around Sgr~A*. We use the available data for S2 to perform a Markov Chain Monte Carlo analysis and find the best-fit estimates for a scalar cloud structure. Our results show no substantial evidence for such structures. When the cloud size is of the order of the size of the orbit of S2, we are able to constrain its mass to be smaller than $0.1\%$ of the central mass, setting a strong bound on the presence of new fields in the galactic centre.
\end{abstract}




\section{Introduction}
The orbit of the star S2 in the Galactic Centre (GC) has been monitored for almost 30 years with both spectroscopic and astrometric measurements, the latter reaching a precision of $\approx 50 \, \mu\rm as$ since the GRAVITY instrument at the Very Large Telescope Interferometer (VLTI) has been put into operation \citep{2017A&A...602A..94G}. S2 is a star with mass around $10-15 \, \mathrm{M}_{\odot}$ orbiting  Sgr~A* with a period of roughly $16$ years and apparent magnitude $K \sim 14$ \citep{Ghez_2003, 2017ApJ...847..120H}. It is part of the so-called Sagittarius A$^*$ cluster, consisting of about 40 stars, known as S-stars, whose orbits are all located within one arcsecond distance from Sgr~A* \citep{Eckart:1996zz, 2002Natur.419..694S, Ghez_2003, 2009ApJ...692.1075G, 2009ApJ...707L.114G, Sabha:2012vc}. The data collected has allowed constraining with unprecedented accuracy both the mass $M$ of the central object and the GC distance $R_0$.  In particular, the trajectory of the S2 star, together with those of other stars in the S-cluster, showed that their motion is determined by a potential generated by a dark object with mass $M \sim 4.3 \cdot 10^6 \mathrm{M}_{\odot}$ at a distance $R_0 \sim 8.3 \, \rm kpc$ \citep{2008ApJ...689.1044G, 2019A&A...625L..10G, GRAVITY:2021xju}, widely believed to be a supermassive black hole  \citep[SMBH,][]{Genzel:2010zy}. This hypothesis has been supported by the direct observations of near-IR flares in the relativistic accretion zone of Sgr~A*, corresponding to the innermost stable circular orbit of a black hole (BH) \citep{2018A&A...618L..10G}, and, most recently, analysing the image of Sgr~A* taken by the Event Horizon Telescope (EHT) which is compatible with the expected appearance of a Kerr BH with such a mass \citep{EventHorizonTelescope:2022wkp}.

While the nature of the central object seems to be well established, its surrounding environment remains mostly unknown. In this context, an especially exciting prospect is that dark matter (DM) may cluster around supermassive BHs, producing spikes in the local density~\citep{Gondolo:1999ef,Sadeghian:2013laa}, leaving imprints in the orbits of stars. The scattering of DM by passing stars or BHs, or accretion by the central BH induced by heating in its vicinities may significantly soften the spike distribution~\citep{Merritt:2002vj,Merritt:2003qk, Bertone:2005hw}. Given the outstanding challenge that DM represents, it is specially important to test the presence of new forms of matter in the GC (for a review
on the GC and how it can be used to constrain DM see \cite{2022arXiv221107008D}). 

Data collected for S2 has been used to test the presence of an extended mass within its apocenter ($r_{\rm apo, S2} = 14 \, \rm mas$) with particular attention to spherically symmetric DM density distributions (see e.g. \cite{Lacroix:2018zmg, Bar:2019pnz, Heissel:2021pcw,GRAVITY:2021xju}).

\cite{Lacroix:2018zmg} used data up to 2016 to fit the size of a DM spike within a halo described by a density profile \citep{1996MNRAS.278..488Z}:
\beq
\rho_{\rm NFW} = \rho_s \left(\frac{r}{r_s} \right)^{-\gamma} \left( 1 + \frac{r}{r_s}\right)^{\gamma -3}\,,
\eeq 
where $r_s$ is the scale radius, $\rho_s$ is the scale density which can be trivially related to the local DM density. 
Lacroix was able to exclude a spike with a radius greater than $10^3$ pc (Figure 2, last plot), which corresponds to $R_{\rm sp} \approx 4.8 \cdot 10^9 \, M$, which can be translated in an upper bound on the total ``environmental'' mass $\delta M$ within the characteristic size of the orbit, $\delta M \lesssim 4-5 \cdot 10^4 \, M_{\odot}$, i.e. $\sim 1\% \, M $. 

\cite{Bar:2019pnz} used similar data to constrain the presence of ultralight dark matter, i.e., matter in the form of a self-gravitating scalar condensate. This assumption fixes the density distribution of the mass profile, and they were able to set an upper bound on the soliton mass of $\delta M \sim 5 \cdot 10^4 \, M_{\odot}$ for a fundamental scalar field with mass $m_s \sim 4 \cdot 10^{-19} $ eV. For $m_s \gtrsim 10^{-18}$ eV the soliton is confined inside S2 periastron and is degenerate with the BH mass. 

\cite{DellaMonica:2022kow} used a similar procedure to derive an upper limit of 
$< 10^{-19} \, \rm eV$ 
on the mass of ultralight boson to beat 95\% confidence level.

Recently, \cite{GRAVITY:2021xju} provided the current $1 \sigma$ upper bound on the environmental mass $\delta M$ within the orbit of S2, namely $\delta M \sim \, 4000 \, M_{\odot}$, or
$0.1 \%$ of the BH mass. This limit was obtained assuming a Plummer model for the matter profile,
\beq
\rho_{\rm Plummer} = \frac{3 f_{\rm PL} M}{4 \pi a_0^3} \left(1 + \left(\frac{r}{a_0}\right)^2\right)^{-5/2} \, ,
\label{plummer_density}
\eeq
with $a_0$ a length scale of the external matter distribution, which has mass $f_{\rm PL}M$. 
In fact, considering a scale length given by roughly S2's apoastron ($a_0 = 0.3 "$), a best-fit value for a fraction of extended mass within S2's orbit of $f_{\rm PL} = (2.7 \pm 3.5) \cdot 10^{-3}$ was found, i.e. $f_{\rm PL}$ is compatible with zero at $1\sigma$ confidence level, and it can be interpreted as a null result. Using, in addition, the orbits of the other four S-stars, upper limits on the extended mass were imposed, of order $10^3 \, M_{\odot}$, equivalent to $0.1 \%$ of the central mass $M$.

Thus far, the profile of the matter distribution has been mostly ad-hoc. Here, we study the possibility that new fundamental fields exist and that they ``condense'' in a bound state around the BH (for a review, see \citet{Brito:2015oca}). These fields might be a significant component of dark matter, or simply as-yet unobserved forms of matter. It is a tantalizing possibility that supermassive BHs might then be used as particle detectors, a possibility that we explore, using the motion of S2 as a probe of the matter content. In this context, the matter profile is known and given by the spatial profile of bound states around spinning BHs~\citep{Detweiler:1980uk,Cardoso:2005vk,Dolan:2007mj,Witek:2012tr,Brito:2015oca}. It can be argued that also in the context of fuzzy dark matter, composed of an ultralight scalar, the near-horizon region is controlled by BH physics, hence governed by the same type of profile we consider here~\citep{Cardoso:2022nzc}. The suggestion that the stars' motion can be used to probe light fields around BHs is not new \citep{Cardoso:2011xi, Ferreira:2017pth, Fujita:2016yav}, but is here explored explicitly with data from the GRAVITY instrument.

\section{The setup}
\label{sec:setup}
Light bosonic fields can arise in a variety of contexts, for example, in string-inspired theories~\citep{Arvanitaki:2009fg}. However, early examples arose out of the need to explain in a natural way the smallness of the neutron electric dipole moment. They invoked the existence of a new axionic, light, degree of freedom~\citep{PhysRevLett.38.1440,PhysRevLett.40.279,PhysRevLett.40.223,Preskill:1982cy,Abbott:1982af,Dine:1982ah}.

In the presence of a spinning BH, small fluctuations of a massive scalar field can be exponentially amplified via superradiance, leading to a condensate -- a bound state -- outside the horizon~\citep{Brito:2015oca}. This structure can carry up to $\sim 10\%$ of the BH mass if grown from vacuum. It is also possible that the scalar soliton existed on its own, for example, if it is part of dark matter, in which case the placing of a BH at its centre will lead to a long-lived structure (a ``cloud'') which on BH scales resembles the superradiant bound states~\citep{Cardoso:2022vpj, Cardoso:2022nzc}.
Here we will be agnostic regarding the origin of the scalar structure, but we will use our knowledge about the spatial profile of bound states around BHs.
\subsection{The scalar field profile}
%
\begin{figure}
\begin{center}
\includegraphics[width=0.45\textwidth]{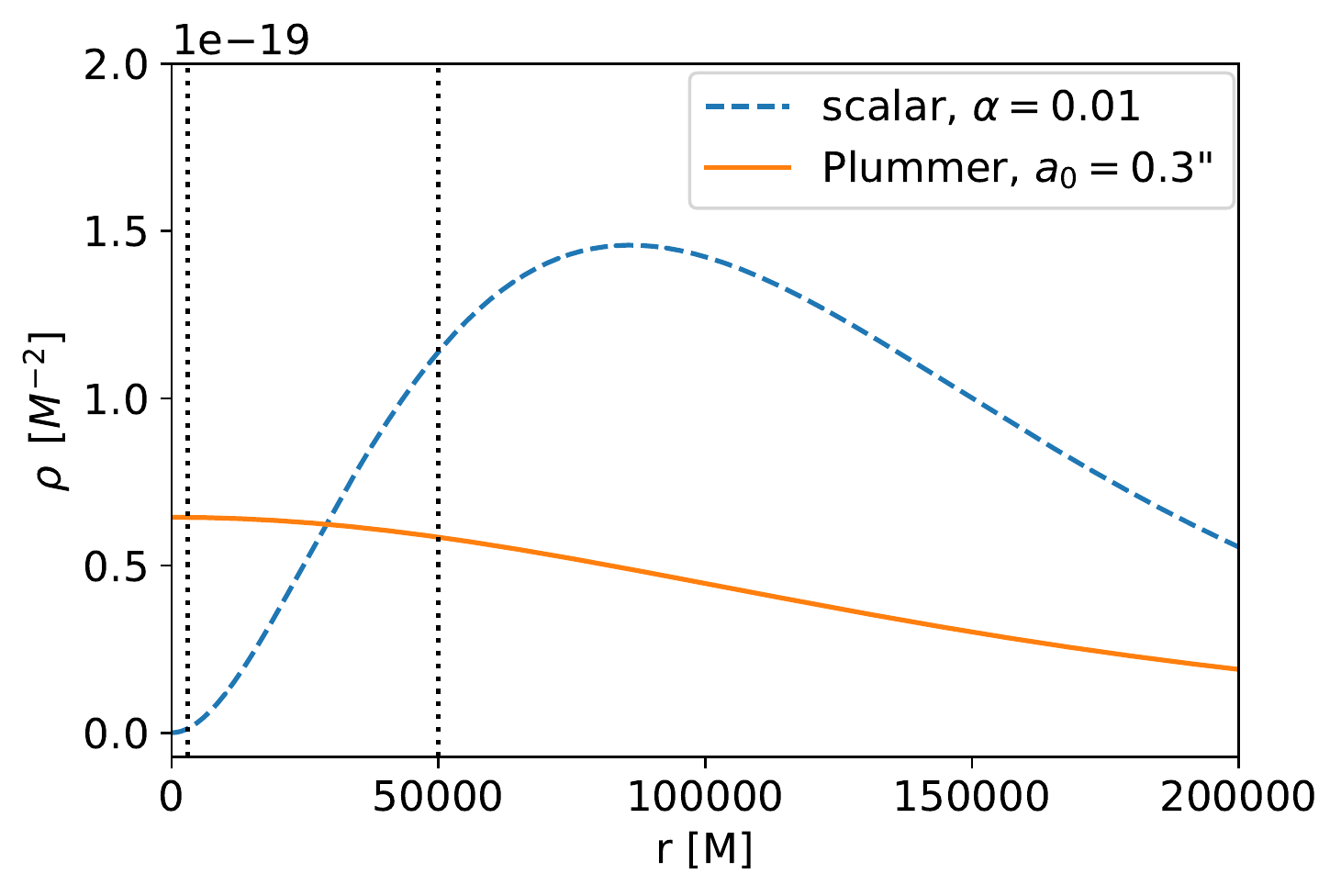} 
\end{center}
\caption{Comparison between the scalar field density in Eq.~\eqref{density} with $\alpha = 0.01$, $\Lambda = 10^{-3}$ and $\theta = \pi/2$ (blue dashed line) and the Plummer density in Eq.~\eqref{plummer_density} with $a_0 = 0.3"$ and $f_{\rm PL} = 10^{-3}$ (orange solid line). Black dotted lines correspond to S2's periastron ($r_{\rm peri} \sim 3000 \, M$) and apoastron ($r_{\rm apo} \sim 50000 \, M$) .
}
\label{fig:comparison_densities}
\end{figure}
Consider a particle moving in a potential given by a central mass $M$ surrounded by a scalar field cloud. Our starting point is the setup developed in \citet{GRAVITY:2019tuf}, and here we recall the most relevant steps of their procedure. 

A system composed of a central BH with mass $M$ and a scalar field minimally coupled to gravity is described by the action
\begin{equation}
S = \int d^4 x \sqrt{-g} \left(\frac{R}{16 \pi G} - \frac{1}{2} g^{\alpha \beta} \psi_{,a}^* \psi_{,b}^* - \frac{\mu^2}{2} \psi \psi^* \right)\,,
\end{equation}
where $R$ is the Ricci scalar, $g_{\mu \nu}$ and $g$ are the metric and its determinant. We assume that the BH spins along the $z-$axis, with adapted spherical coordinates $(t, r, \theta, \phi)$, with $\theta=\pi/2$ defining the equator. The scalar $\psi(t, r, \theta, \phi)$ is a complex field, and $\mu$ is a mass parameter for the scalar field. It is related to the physical mass $m_s$ via $\mu = m_s c/ \hbar$ and to the (reduced) Compton wavelength of the particle via $\lambdabar_C = \mu^{-1}$. The principle of least action results in the Einstein-Klein-Gordon system of equations, where the energy-momentum tensor of the scalar field can be written as
\begin{equation}
T_{\mu \nu} = \frac{1}{2} \left[ \psi_{, \mu} \psi^*_{, \nu} + \psi_{, \nu} \psi^*_{, \mu} - g_{\mu \nu} \left( \psi^{, \sigma} \psi^*_{, \sigma} + \mu^2 | \psi|^2 \right) \right] \, .
\end{equation}
In the low-energy limit, i.e. neglecting terms of $\mathcal{O}(c^{-4})$, the energy density of the field reads
\begin{equation}
\rho =  \frac{m_s^2 c^2}{\hbar^2} |\psi |^2  = \mu^2 |\psi|^2 = \left(\frac{\alpha}{M}\right)^2 |\psi|^2 \,,
\label{density}
\end{equation}
where we have defined the dimensionless mass coupling $\alpha$ as
\beq
\alpha = \left[\frac{G M}{c^2} \right] \left[\frac{m_s c}{\hbar} \right]
\label{alpha}
\eeq
From now on we will use natural units ($G = c = \hbar = 1$) unless otherwise stated.

The solution of the Klein-Gordon equation for the field $\psi$ on a Kerr background can be decomposed into a radial and an angular part, as $\psi = e^{-i \omega t + i m \phi} S_{lm} (\theta) R_{l m}(r)$, where $l, m$ are the angular modes, and $\omega \sim \mu$ defines the frequency of the field. In the limit of small coupling ($\alpha \ll 1$), the radial part is proportional to the generalised Laguerre polynomials $L_{n}^{2l +1}$ and the angular part becomes $S_{lm}(\theta) = P_l^m (\cos\theta)$ with $ P_l^m (\cos\theta)$ being the associated Legendre polynomials. In this approximation, the fundamental mode $n=0$, $l=m=1$ of the scalar field is given by \citep{Brito:2014wla}
\beq
 \psi = A_0 e^{-i(\omega t - \phi)} \frac{r}{M} \alpha^2 e^{-\frac{r \alpha^2}{ M}} \sin \theta\,,
 \label{scalar_profile}
\eeq
where the amplitude of the field $A_0$ is related to the mass of the cloud via
\beq
M_{\rm cloud} = \int \rho s^2 \sin \theta d \theta ds d \phi = \frac{64 \pi A_0^2}{\alpha^4} M \, .
\eeq
%
We can now use the energy density of the field to solve Poisson's equation $\nabla^2 U_{\rm sca} =  4 \pi \rho$, using the usual harmonic decomposition implemented in \citet{PoissonWill2012}, i.e., expanding all quantities in spherical harmonics $Y_{lm}=Y_{lm}(\theta,\phi)$. For the energy density computed in \eqref{density} the only non-zero terms that contribute to the scalar potential are the $l = m = 0$ and $l = 2$, $m=0$ terms, resulting in a potential given by
\beq
\begin{split}
U_{\rm sca}  & =  4 \pi \left[ \frac{q_{00}}{r} Y_{00} + p_{00} Y_{00} \right] + \frac{4 \pi}{5} \left[ \frac{q_{20}}{r^3} Y_{20} + p_{20} r^2 Y_{20} \right] \\
& =   \Lambda \left( P_1(r) + P_2(r) \cos^2 \theta \right)\,,
\end{split}
\label{scalar_potential}
\eeq
where $\Lambda = M_{\rm cloud}/M$ is the fractional mass of the scalar field cloud to the BH mass, 
\beq
\begin{split}
P_1(r) & = \frac{M}{r} + \frac{3 M^3}{r^3 \alpha^4} - \frac{e^{- \frac{r \alpha^2}{M}}}{16 M^2 r^3 \alpha^4} \left( 48 M^5 + 48 M^4 r \alpha^2 + 40 M^3 r^2 \alpha^4 \right. \\
& \left. + 20 M^2 r^3 \alpha^6 + 6 M r^4 \alpha^8 + r^5 \alpha^{10} \right) \, ,
\end{split}
\eeq
and 
\beq
\begin{split}
P_2(r) & = - \frac{9 M^3}{ r^3 \alpha^4} + e^{-\frac{r \alpha^2}{M}} \left(\frac{9M}{2 r} + \frac{9 M^3}{r^3 \alpha^4} + \frac{9 M^2}{r^2 \alpha^2} + \frac{3\alpha^2}{2} \right. \\
& \left. + \frac{3 r \alpha^4}{8 M} + \frac{r^2 \alpha^6}{16 M^2} \right) \, .
\end{split}
\eeq

In Figure~\ref{fig:comparison_densities} we show the difference between the scalar field density in \eqref{density} along the equator ($\theta = \pi/2$, with $\Lambda = 10^{-3}$ 
and $\alpha = 0.01$) and the density given by a Plummer profile \eqref{plummer_density}, where we use the same values as in~\citet{GRAVITY:2021xju}: $a_0 = 0.3\,\arcsec$ and $f_{\rm PL} = 10^{-3}$. 

\cite{GRAVITY:2019tuf} showed that a scalar field cloud described by the potential \eqref{scalar_potential} can leave imprints in the orbital elements of S2 if its mass coupling constant is in the range
\beq
0.005 \lesssim \alpha \lesssim 0.05 \, ,
\label{range_alpha}
\eeq 
assuming a fixed direction of the BH spin axis with respect to the plane of the sky, which corresponds to an effective mass of the field in the range $10^{-20} \, \rm eV \lesssim \mu \lesssim 10^{-18} \, \rm eV$. However, \cite{Kodama:2011zc} showed that for an SMBH with the mass of Sgr~A*, the allowed range of effective masses that can engage a superradiant instability on a timescale smaller than the cosmic age is $10^{-18} \, \rm eV \lesssim \mu \lesssim 10^{-15} \, \rm eV$. Hence, if a cloud exists and leaves detectable imprints in the orbit of S2, then its formation and existence must be explained by means of a different physical process, as discussed in Sec.~\ref{sec:setup}. However, since the variations in the orbital elements induced by the cloud are potentially detectable with the current precision of the GRAVITY instrument, it is worth comparing these theoretical expectations with the available data. In particular we are interested in fitting the fractional mass of the cloud $\Lambda = M_{\rm cloud}/M$ for a fixed value of the mass coupling constant $\alpha$.
\subsection{The equations of motion}
To obtain the equations of motion of a particle moving in a central potential plus the toroidal scalar field distribution described by \eqref{scalar_profile} we started from the Lagrangian
\beq
\mathfrak{L} = \frac{1}{2} \left( \dot{r}^2 + r^2 \dot{\theta}^2 + r^2 \sin^2 \theta \dot{\phi}^2 \right) + U(r, \theta) \, ,
\label{lagrangian}
\eeq
where
\beq
U(r, \theta) = \frac{M}{r} + \Lambda \left(P_1(r) + P_2(r) \cos^2 \theta \right)\,,
\label{total_potential}
\eeq
is the sum of the Newtonian and the scalar potential. Solving the Euler-Lagrange equations translates into having the following equations of motion,
\beq
\begin{split}
& \ddot{r} = - \frac{M}{r^2} + r \left(\dot{\theta}^2 + \sin^2 \theta \dot{\phi}^2 \right) + \Lambda \left(P_1'(r) + P_2'(r) \cos^2 \theta \right) \\
& \ddot{\theta} = \cos \theta \sin \theta \dot{\phi}^2 - \frac{2}{r} \dot{r} \dot{\theta} - \frac{\Lambda P_2 (r) \sin 2\theta}{r^2}  \\
&  \ddot{\phi} = - \frac{2 \dot{\phi}}{r} \left( \dot{r} + \cot \theta \, r \dot{\theta} \right)
\end{split}\,,
\label{true_eom}
\eeq
where the prime (dot) indicates a derivative with respect to the radial (time) coordinate. Since the Schwarzschild precession has been detected in the orbit of S2 at $7 \sigma$ confidence level \citep{GRAVITY:2021xju}, we also included the first Post Newtonian correction in the equations of motion. The acceleration term is given by \citep{Will:2007pp} 
\beq
\boldsymbol{a}_{1 \rm PN} = f_{\rm SP} \frac{ M}{r^2} \left[\left(\frac{4M}{r} - v^2\right) \frac{\boldsymbol{r}}{r} + 4 \dot{r}\boldsymbol{v}  \right] \, ,
\eeq 
where $\boldsymbol{r} = r \hat{r}$, 
\beq
\boldsymbol{v} = \left(\dot{r} \hat{r}, r \dot{\theta}\hat{\theta}, r \dot{\phi} \sin \theta \hat{\phi} \right) \, ,
\eeq
and $ v = |\boldsymbol{v}|$. Here we have also introduced the dimensionless parameter $f_{\rm SP}$ that quantifies the Schwarzschild precession, and it is found to be $f_{\rm SP} = 0.99 \pm 0.15$ \citep{GRAVITY:2021xju}. In this work we fixed $f_{\rm SP} = 1$.

If we impose $\Lambda = 0$ and $f_{\rm SP} = 0$ we recover the classical motion of a particle orbiting a central point mass. The 6 initial conditions for the set of equations in \eqref{true_eom} can be obtained from the analytical solution of the Keplerian two-body problem, namely
\beq
\begin{split}
    & r_0 = \frac{a_{\rm sma}(1 - e^2)}{1 + e \cos \phi_0}\, , \,\,\,\,\,\,\,\,\,\,\,\,\,\,\,\,\,\,\,\,\,\,\,\,\,\,\,\,\,\,\,\,\, \dot{r}_0 =  \frac{2 \pi e a_{\rm sma} \sin \mathcal{E}}{P(1 - e \cos \mathcal{E})} \\
    & \theta_0 = \frac{\pi}{2} \, , \,\,\,\,\,\,\,\,\,\,\,\,\,\,\,\,\,\,\,\,\,\,\,\,\,\,\,\,\,\,\,\,\,\,\,\,\,\,\,\,\,\,\,\,\,\,\,\,\,\,\,\,\,\,\,\,\,\,\,\,\, \dot{\theta} = 0 \\
    & \phi_0 = 2 \arctan\left(\sqrt{\frac{1 + e}{1 - e}} \tan \frac{\mathcal{E}}{2} \right) \, , \,\,\,\, \dot{\phi}_0 = \frac{2 \pi (1-e)}{P(e \cos \mathcal{E} - 1)^2} \sqrt{\frac{1 + e}{1 - e}}
\end{split}
\label{initial_cond}
\eeq
where $e, a_{\rm sma}, P$ are the eccentricity, the semi-major axis and the period of the orbit, respectively, while $\mathcal{E}$ is the eccentric anomaly evaluated from Kepler's equation: $\mathcal{E} - e \sin \mathcal{E} - \mathcal{M} = 0$, where $\mathcal{M} = n (t - t_p)$ is the mean anomaly, $n = 2 \pi/P$ is the mean angular velocity and $t_p$ is the time of periastron passage. Details about how we performed the numerical integration and how we solved Kepler's equation are reported in Appendix \ref{app:num_integration}. The solution of the previous equations of motion gives the spherical coordinates of the star in the BH reference frame, related with Cartesian coordinates $\{x_{\rm BH}, y_{\rm BH}, z_{\rm BH} \}$ via the usual transformation. In this frame, $z_{\rm BH}$ is aligned with the BH spin axis. Following \citet{Grould:2017bsw} we can define a new reference frame $\{x', y', z_{\rm obs}\}$ such that $x' = \rm DEC$, $y' = \rm R.A.$ are the collected astrometric data, $z_{\rm obs}$ points towards the BH and $v_{z_{\rm obs}}$ corresponds to the radial velocity. 
Despite most of the S2 motion occurring in a Newtonian regime (i.e. with $ v \ll 1$) making the above classical approximation appropriate, near the periastron it reaches a total space velocity of $v \approx 7650 \, \rm km/s \sim 10^{-2}$. 
In this region the numerical solution $v_{z_{\rm obs}}$ obtained from Eqs.~\eqref{true_eom} must be corrected. We include the two main relativistic effects in order to model the measured radial velocity $V_R$: the relativistic Doppler shift and the gravitational redshift. 
Moreover, due to the finite speed of light propagation, the dates of observation $t_{\rm obs}$ are generally different from the dates of emission $t_{\rm em}$. This is a pure classical effect known as R{\o}mer's delay, and for S2 we have $\Delta t = t_{\rm em} - t_{\rm obs} \approx 8 \, \rm days$ on average over the entire orbit. Including this effect in our simulation requires solving the so-called R{\o}mer's equation, namely:
\begin{equation}
t_{\rm obs} - t_{\rm em} - z_{\rm obs}(t_{\rm em}) = 0
\label{roemer_equation}
\end{equation}
(here we corrected a minus sign in \citet{Grould:2017bsw}) that we solved using its first-order Taylor's expansion, as already done in \citet{GRAVITY:2018ofz, Heissel:2021pcw}. 

Details about how to implement the transformation between the orbital frame and the observer frame, how to include the relativistic corrections and how we solved Eq.~\eqref{roemer_equation} are reported in Appendix \ref{app:relativistic_effects}.

\subsection{Data}
The set of available data $D$ can be divided as follows: 
\begin{itemize}
\item[a)] Astrometric data $\rm DEC$, $\rm R.A.$ 
    \begin{itemize}
        \item 128 data points collected using both the SHARP camera at New Technology Telescope (TNN) between 1992 and 2002 ($\sim$ 10 data points, accuracy $\approx 4 \, \rm mas$) and the NACO imager at the VLT between 2002 and 2019 (118 data points, accuracy $\approx 0.5 \, \rm mas$);
    
        \item 76 data points collected by GRAVITY at VLT between 2016 and April 2022 (accuracy $\approx 50 \, \rm \mu as$).
    \end{itemize}
    
\item[b)] Spectroscopic data $V_R$
    \begin{itemize}
        \item 102 data points collected by SINFONI at the VLT (100 points) and NIRC2 at Keck (2 points) collected between 2000 and March 2022 (accuracy in good conditions $\approx 10-15 \, \rm km/s $).
    \end{itemize}
\end{itemize}

\subsection{Model fitting approach}

To fit S2 data we perform a Markov Chain Monte Carlo (MCMC) analysis using the Python package  \textsc{emcee} \citep{2013PASP..125..306F}. 
The fitting procedure is as follows: we set the value of the mass coupling $\alpha$ roughly within the range reported in~\eqref{range_alpha}. For any given value of $\alpha$ we fit for the following set of parameters,
\begin{equation}
    \Theta_i = \{e, a_{\rm sma}, \Omega_{\rm orb}, i_{\rm orb}, \omega_{\rm orb}, t_p, R_0, M, x_0, y_0, v_{x_0}, v_{y_0}, v_{z_0}, \Lambda \}\,,
    \label{emcee_parameters}
\end{equation}
where $\Omega_{\rm orb}$, $i_{\rm orb}$ and $\omega_{\rm orb}$ are the three angles used to project the orbital frame in the observer reference frame using the procedure reported in Appendix \ref{app:coord_transf}. The additional parameters $\{x_0, y_0,v_{x_0}, v_{y_0}, v_{z_0} \}$ characterise the NACO/SINFONI data reference frame with respect to Sgr~A* \citep{2015MNRAS.453.3234P}. 
The log-likelihood is given by
\begin{equation}
    \ln \mathcal{L} = \ln \mathcal{L}_{\rm pos} + \ln \mathcal{L}_{\rm vel}\,,
\end{equation}
where 
\begin{equation}
    \ln \mathcal{L}_{\rm pos} = - \sum_{i=1}^{N} \left[ \frac{ (\rm DEC_{i} - \rm DEC_{\rm model, i})^2}{\sigma_{\rm DEC_{i}}^2}  +  \frac{ (\rm R.A._{i} - \rm R.A._{\rm model, i})^2}{\sigma_{\rm R.A._{i}}^2} \right]\,,
\end{equation}
and 
\begin{equation}
    \ln \mathcal{L}_{\rm vel} = - \sum_{i=1}^{N} \frac{ (V_{R, i} - V_{\rm model, i})^2}{\sigma_{V_{R, i}}^2} \, .
\end{equation}
The priors we used are listed in Table~\ref{table:priors}. We used uniform priors for the physical parameters, i.e. we only imposed physically motivated bounds and Gaussian priors for the additional parameters describing NACO data, since the latter have been instead well constrained by previous work by \citet{2015MNRAS.453.3234P} and are not expected to change.
\begin{table}
\caption{Uniform priors used in the MCMC analysis. Initial guesses $\Theta_i^0$ coincide with the best-fit parameters found by \textbf{minimize}.} 
\label{table:priors}
\begin{tabular}{lccc}
    \hline
    Parameter & $\Theta_i^0$ & Lower bound & Upper bound \\
    \hline
    $e$ & 0.88441 & 0.83 & 0.93 \\[2pt] 
    $a_{\rm sma}$ [as] & 0.12497 & 0.119 & 0.132 \\[2pt]  
    $i_{\rm orb} \, [^\circ] $  & $134.69241$ & 100 & $150$ \\ [2pt] 
    $\omega_{\rm orb} \, [^\circ]$ & $66.28411$ & 40 & $90$ \\ [2pt] 
    $\Omega_{\rm orb} \, [^\circ]$ & $228.19245$ & $200$ & $250$ \\ [2pt] 
    $t_p $ [yr] & 2018.37902 & 2018 & 2019 \\ [2pt] 
    $M \, [10^6 \, M_{\odot}] $ & $4.29950$ & 4.1 & 4.8\\[2pt] 
    $ R_0 \, \rm [10^3 \, pc]$ & 8.27795 & 8.1 & 8.9\\ [2pt] 
    $\Lambda$ & 0.001 & 0 & 1 \\
    \hline
\end{tabular}
\end{table}
\begin{table}
\caption{Gaussian priors used in the MCMC analysis. Initial guesses $\Theta_i^0$ coincide with the best-fit parameters found by \textbf{minimize}. $\xi$ and $\sigma$ represent the mean and the standard deviation of the distributions, respectively, and they come from \citet{2015MNRAS.453.3234P}.} 
\begin{tabular}{lccc}
    \hline
    Parameter & $\Theta_i^0$ & $\xi$ & $\sigma$ \\
    \hline
    $ x_0 \, \rm [mas]$ & -0.244 & -0.055 & 0.25 \\ [2pt] 
    $ y_0 \, \rm [mas]$ & -0.618 & -0.570 & 0.15 \\ [2pt] 
    $ v_{x_0} \, \rm [mas/yr]$ & 0.059 & 0.063 & 0.0066 \\ [2pt] 
    $ v_{y_0} \, \rm [mas/yr]$ & 0.074 & 0.032 & 0.019 \\[2pt] 
    $ v_{z_0} \, \rm [km/s]$ & -2.455 & 0 & 5 \\[2pt] 
    \hline
\end{tabular}
\end{table}
The initial points $\Theta_i^0$ in the MCMC are chosen such that they minimise the $\chi^2$ when $f_{\rm SP} = 1$ and $\Lambda = 0$. The minimisation is performed using the Python package \textbf{lmfit.minimize} \citep{lmfit2014} with Levenberg-Marquardt method. In the sampling phase of the MCMC implementation, we used 64 walkers and $10^5$ iterations. Since we started our MCMC at the minimum found by \textbf{minimize} we skipped the burning-in phase and we used the last $80\%$ of the chains to compute the mean and standard deviation of the posterior distributions. The convergence of the MCMC analysis is assured by means of the auto-correlation time $\tau_c$, i.e. we ran $N$ iterations such that $N \gg 50 \, \tau_c$.

\begin{figure*}
\includegraphics[width=\textwidth]{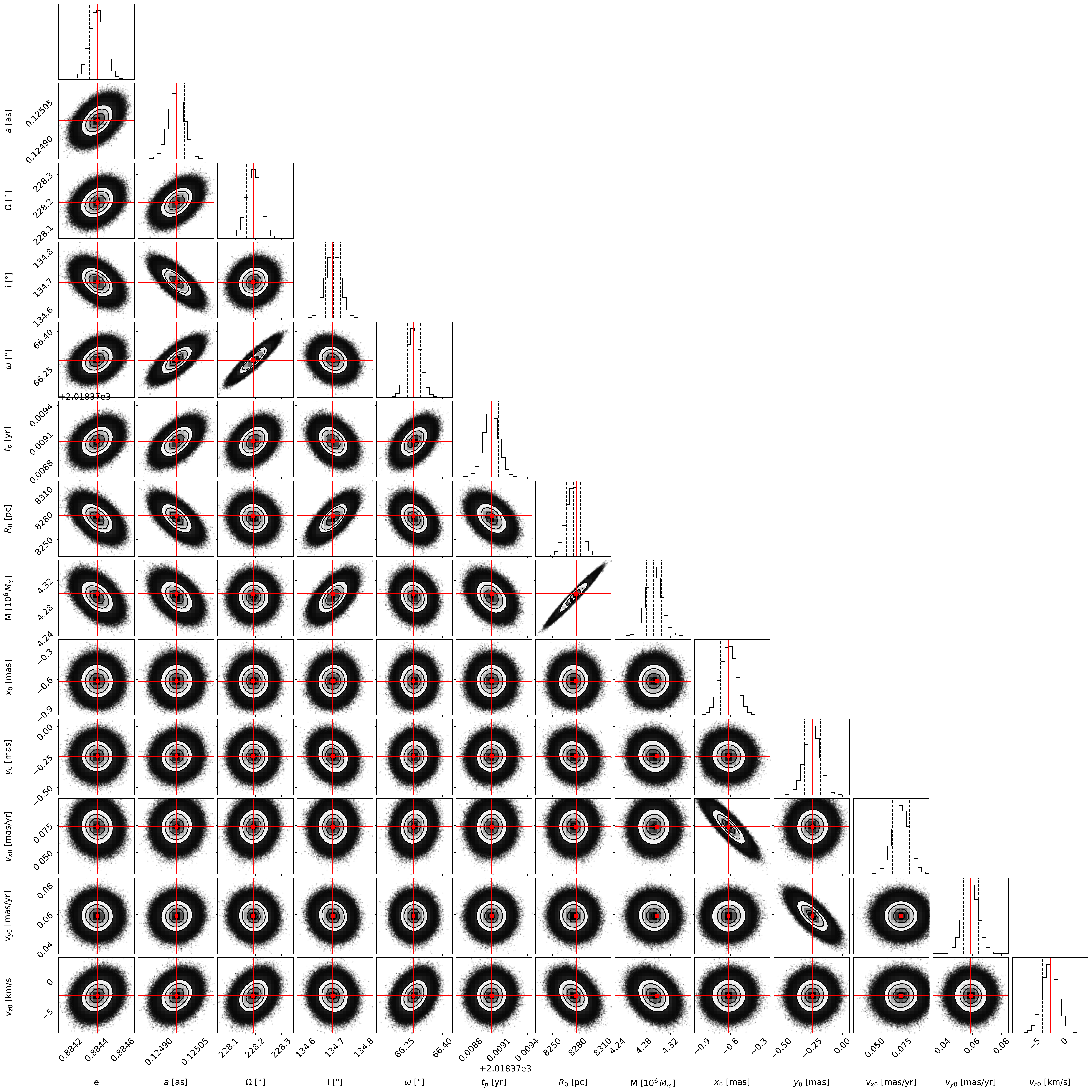}
    \caption{Corner plot of the fitted parameters with $f_{\rm SP} = 1$ and $\Lambda = 0$. Red lines represent values from \textbf{minimize}, while dashed black lines represent the mean value and $1\sigma$ interval of the posterior distributions.}
    \label{fig:corner_plot_1PN}
\end{figure*}

In a first preliminary check we set $\Lambda = 0$ and we fit for the first 13 parameters of ~\eqref{emcee_parameters} imposing $f_{\rm SP} = 1$. In Figure~\ref{fig:corner_plot_1PN} we report the corner plot of the parameters, which are in very good agreement with the previous best estimates obtained in \citet{GRAVITY:2021xju}.
In the following, we assume that $z_{\rm BH}$ is aligned with $z_{\rm orb}$, i.e. the direction of the BH spin axis is aligned with the angular momentum of the S2 orbit. This means that the motion happens in the equatorial plane ($\theta = \pi/2$) of the BH and the initial conditions for the numerical integration of the orbit are those reported in~\eqref{initial_cond}. 
We fit for the 14 parameters listed in~\eqref{emcee_parameters}.

\section{Results}
\begin{figure*}
\centering
\includegraphics[width=0.7\textwidth]
{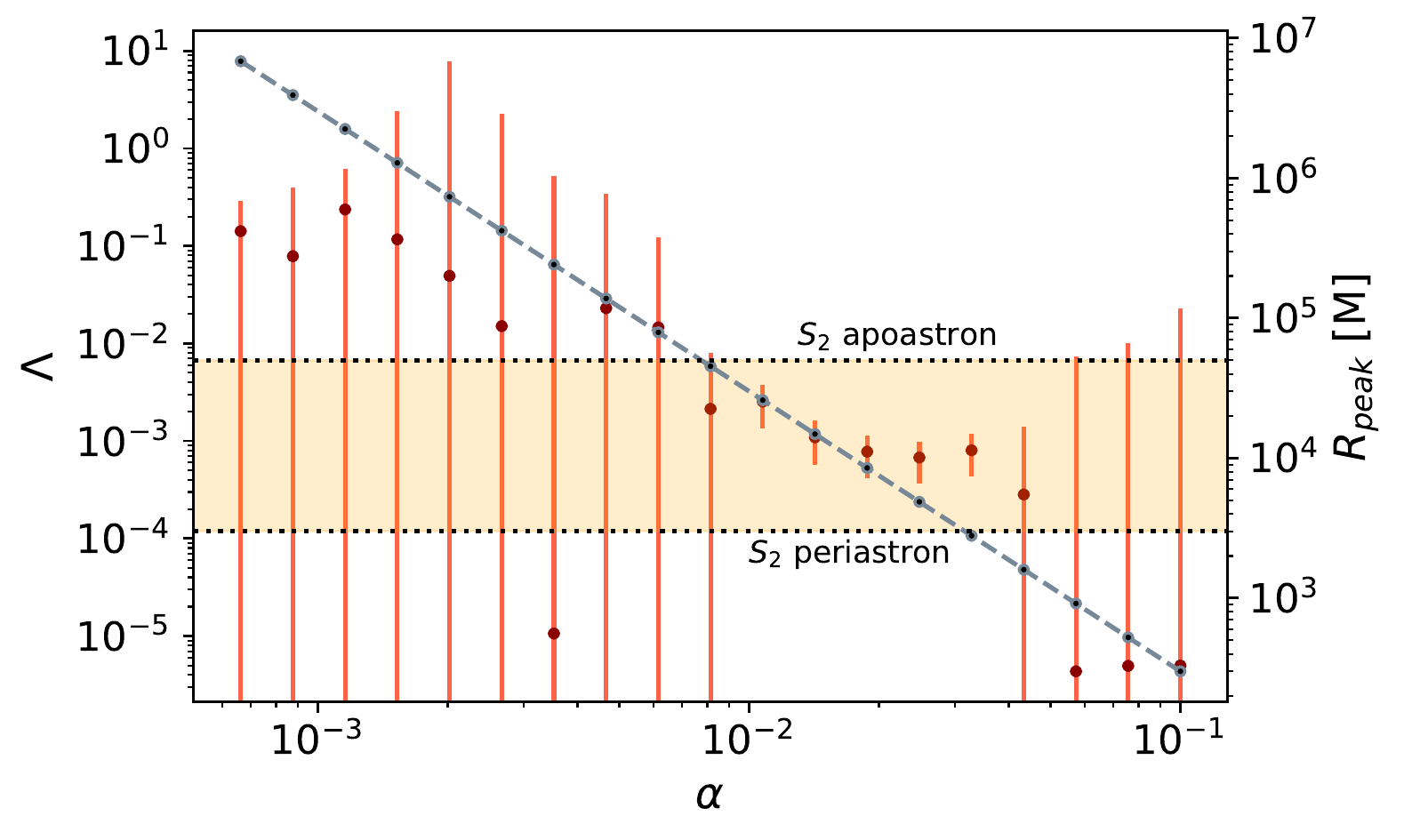}
\caption{Best fit values of $\Lambda$ with 1$\sigma$ uncertainty when $\alpha$ is fixed and it is varied over the range $[6 \cdot 10^{-4}, 10^{-1}]$. The dashed grey line represents $R_{\rm peak}$ as a function of $\alpha$ as illustrated in \eqref{r_peak}. The yellow band represents the orbital range of S2 delimited by its apoastron and periastron positions. Although a nonzero value of $\Lambda$ is apparent for a restricted range of $\alpha$, the statistical significance of this finding is not significant, see Table~\ref{best_fit_lambda}.}
\label{fig:chi_squared_analysis}
\end{figure*}

Before running the MCMC algorithm we used a $\chi^2$ minimiser to evaluate the best-fit values of $\Lambda$ 
and to quantify how accurately we can constrain the scalar cloud mass. Results are summarised in Fig.~\ref{fig:chi_squared_analysis}. For very small ($\alpha \lesssim 0.0035$) or large ( $\alpha \gtrsim 0.045$) values of $\alpha$, $\Lambda$ has very large uncertainties, and the results are compatible with $\Lambda=0$, i.e., having a vacuum environment.

Uncertainties on $\Lambda$ become much smaller in the range $0.01 \lesssim \alpha \lesssim 0.03$. The underlying reason for this can be understood from the effective peak position of the scalar density distribution
\beq
R_{\rm peak} = \frac{\int_0^{\infty} \rho \bar{r} d\bar{r}}{\int_0^{\infty} \rho d\bar{r}} = \frac{3 M}{\alpha^2} \,.
\label{r_peak}
\eeq
For the range of $\alpha$ above, one finds $3000 M \lesssim R_{\rm peak} \lesssim 30000 \, M$, i.e. when $R_{\rm peak}$ is located between S2's apoastron and periastron and the star crosses regions of higher density. This analysis is reported in Fig.~\ref{fig:chi_squared_analysis}, where we show the behaviour of $R_{\rm peak}$ as a function of $\alpha$, dictated by Eq.~\eqref{r_peak}, and S2's apoastron and periastron.

Notice that Fig.~\ref{fig:chi_squared_analysis} seems to indicate that the motion of S2 is compatible with a cloud of scalar field for $0.01 < \alpha < 0.03$. However, as we now discuss, the statistical evidence for a nonzero $\Lambda$ is not significant.

\begin{figure*}
\includegraphics[width=\textwidth]{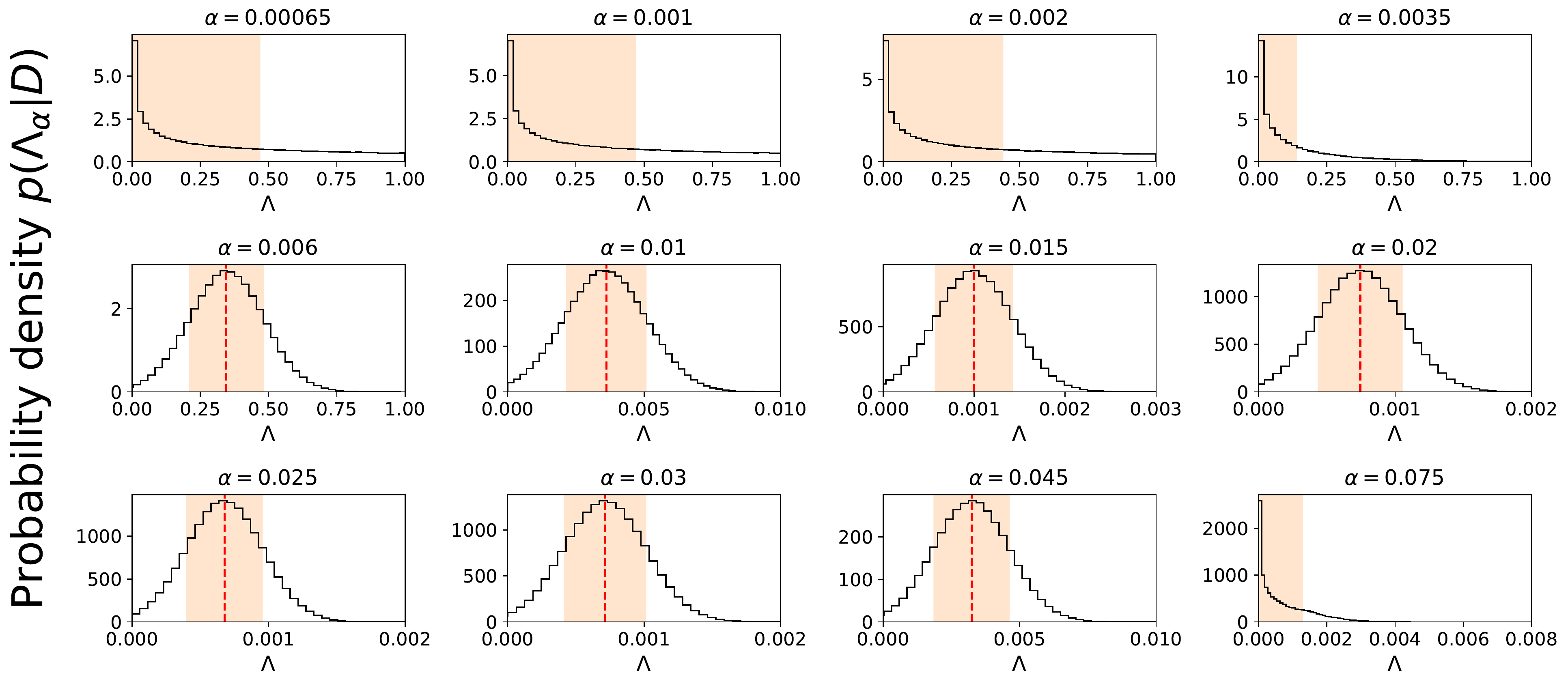}
    \caption{Posterior probability densities $P(\Lambda_{\alpha}|D)$ for different values of $\alpha$. Red dashed lines represent the mean value of  Gaussian distributions (which coincides with the MLE $\hat{\Lambda}$), while orange bands correspond to $1 \sigma$ confidence level, i.e. $\approx 68\%$ of $P(\Lambda_{\alpha}|D)$ lies in that region.}
    \label{fig:lambdas}
\end{figure*}
MCMC results confirm the trend observed in Fig.~\ref{fig:chi_squared_analysis} but provide more insight into how $\Lambda$ is distributed in the range of $\alpha$ considered. In particular, we looked for the maximum likelihood estimator (MLE) of $\Lambda$, i.e.  $\hat{\Lambda} = \rm arg \, max \, \mathcal{L}(\Lambda; D)$. Results are summarised in Fig.~\ref{fig:lambdas}.
For $0.006 < \alpha < 0.075$ the posteriors $P(\Lambda_{\alpha}|D)$ look like normal distributions. Here $\hat{\Lambda}$ and associated uncertainties coincide with the mean and standard deviation of the distributions, and they are roughly the same reported in Fig.~\ref{fig:chi_squared_analysis}. However, when we move away from this range, the posteriors start to be peaked around zero and $\hat{\Lambda}$ does not coincide with the mean value of the distributions anymore, as a result of the prior bounds we imposed on $\Lambda$.  
Since in these cases $\hat{\Lambda}$ is always very close to zero (and far below the precision of current instruments), we estimated $\Lambda_1$ and $\Lambda_2$ such that $P(\Lambda_{\alpha} < \Lambda_1|D) \approx 68 \%$ and $P(\Lambda_{\alpha} < \Lambda_2|D) \approx 99 \%$ of $P(\Lambda_{\alpha}|D)$.
In this way we were able to obtain a rough upper bound on the fractional mass at $1,\,3 \sigma$ confidence levels, reported in parenthesis in Table~\ref{best_fit_lambda}. We notice also that for smaller values of $\alpha$, $P(\Lambda_{\alpha}|D)$ flattens out, showing the difficulties of finding a meaningful MLE $\hat{\Lambda}$ as soon as the cloud is located far away from S$2$'s apoastron. 
These features are shown in Figure~\ref{fig:lambdas}, where we report the one-dimensional projection of the (marginalised) posterior distributions of $\Lambda$ for the values of $\alpha$ reported in Table~\ref{best_fit_lambda}. 
We also show the mean (red dashed line) when distributions are normal and the $1 \sigma$ confidence interval (orange band, evaluated as explained above when the distribution is non-normal). Not surprisingly, we noticed that basically no relevant information can be extracted from those confidence intervals when $R_{\rm peak}$ is far from S$2$'s apoastron. However, in the case with $\alpha = 0.075$, which corresponds to $R_{\rm peak} \approx 530 \, M$, we found that $\Lambda \lesssim 5 \cdot 10^{-3}$ at 3 $\sigma$ confidence level, roughly recovering the upper bound $\delta M \lesssim 10^{-3} \, M$ found in \citet{GRAVITY:2021xju}.

In order to determine the statistical significance of our results we computed the Bayes factor $K$, i.e. the ratio of the maximum likelihood computed for different values of $\alpha$ and $\hat{\Lambda}$ reported in Table~\ref{best_fit_lambda} (that we call model $\alpha$) to the maximum likelihood associated with the non-perturbative case (model $0$). According to~\citet{Kass:1995loi} if $1 \leq \log_{10} K \leq 2$ there is a strong evidence that model $\alpha$ is preferred over model $0$, while if $\log_{10} K > 2$ the strength of evidence is decisive. Negative values of $\log_{10} K$ correspond to negative evidence, i.e. model $0$ is preferred over model $\alpha$.  
As expected, we found $\log_{10} K \ll 1$ every time the cloud is located far away from S$2$ orbital range. In contrast, when $r_{\rm apo, S2} \lesssim R_{\rm peak} \lesssim r_{\rm peri, S2}$ there is only mild evidence that model $\alpha$ is preferred over model $0$ (we found $\log_{10} K < 2$ always). 

\begin{table}
\caption{Maximum Likelihood Estimator $\hat{\Lambda}$ with associated $1 \sigma$ error and Bayes factors $\log_{10} K$ for different values of $\alpha$. The measurements for each $\alpha$ are not independent (the same orbit was used to derive them) and therefore cannot be combined to derive a more stringent upper limit.
For non-normal distributions we report $\Lambda_1$ and $\Lambda_2$ defined such that $P(\Lambda_{\alpha}<\Lambda_1|D) \approx 68\%$ and $P(\Lambda_{\alpha}<\Lambda_2|D) \approx 99\%$ of $P(\Lambda_{\alpha}|D)$.} 
\begin{tabular}{lll}
\hline
$\alpha$ & $\hat{\Lambda}$ & $\log_{10} K$  \\
\hline
    $ 0.00065$ & $\lesssim (0.470 \,, 0.980)$  & 0.09 \\ [0.5pt] \hline
    $ 0.001$ & $\lesssim (0.470 \,, 0.980)$ & 0.08 \\ \hline
    $ 0.002$ & $\lesssim (0.440 \,, 0.978)$ & -0.06 \\ \hline
    $ 0.0035$ & $\lesssim (0.140 \,, 0.780)$ & -10.58 \\ \hline
    $ 0.006$ &  $0.34671 \pm 0.13666$ & 1.44  \\ \hline
    $ 0.01$ & $0.00361 \pm 0.00147$ & 1.29  \\ \hline
    $ 0.015$ & $0.00101 \pm 0.00042$ & 1.24  \\ \hline 
    $ 0.02$ &  $0.00075 \pm 0.00030$ & 1.33 \\ \hline 
    $ 0.025$ & $0.00068\pm 0.00028$ & 1.35 \\ \hline 
    $ 0.03$ & $0.00073 \pm 0.00029$ & 1.33  \\ \hline 
    $ 0.045$ &  $0.00328\pm 0.00135$ & 1.27 \\  \hline 
    $ 0.075$ & $\lesssim (0.0013 \,, 0.0052)$  & 0.0001 \\ 
    \hline
\end{tabular}
\label{best_fit_lambda}
\end{table}
%





\section{Discussion}
Precision observations by the GRAVITY instrument can now be used to set exquisite constraints on possible dark matter structures around Sgr~A*. We have shown that with current observations, scalar clouds -- possibly of superradiant origin, with mass couplings in the range $\alpha \in [0.015, 0.045]$ can be ruled out, for cloud masses $\Lambda\gtrsim 0.1\%$ of the central BH mass (equivalent to $\delta M \sim \, 4000 \, M_{\odot}$). It is similar to that of \cite{GRAVITY:2021xju}, who provided a $1 \sigma$ upper bound of $0.1 \%$ of $M$ on the observational dark mass within the orbit of S2 assuming a Plummer profile for the distribution.  

We also note that, for certain scalar couplings $\alpha$, observational data are well fitted by a non-zero value of $\Lambda$ of order $10^{-3}$. However, all these values of $\Lambda$ are consistent with zero within the $3 \, \sigma$ confidence interval. The computation of the Bayes' factor showed that this perturbed model is only mildly preferred over the non-perturbed model predicting a single central BH without a cloud. We conclude that there is no strong evidence to claim the existence of a scalar cloud around Sgr~A* described by our setup.

Stronger constraints -- or a detection -- require more observations or the inclusion of other stars of the S-cluster in the fit. However, since the potential describing the cloud is non-spherically symmetric, the inclination of stars with each other plays a fundamental role - at least in theory - and this same analysis can not be performed straightforwardly. For the same reason, we were forced to set an initial angular position for S2 co-planar with the BH equator ($\theta = \pi/2$). This is the simplest choice but also the one that maximises the scalar potential in Eq.~\eqref{scalar_potential}, i.e. our chances to actually detect the cloud. 
We can try to quantify the error we are making in setting the initial angular position of the star, by looking at the difference in the orbits for two different initial inclinations: $\theta = \pi/2$ and $\theta = 0$, focusing on the interesting range of $\alpha$: $0.01 \leq \alpha \leq 0.045$. We found that the maximum relative (percentage) difference in the astrometry is achieved for $\alpha = 0.01$, where $\Delta \rm DEC \sim \Delta \rm R.A. \approx 25\%$, while the maximum difference in the radial velocity is found to be $\Delta V_R \approx 15\% $ for $\alpha = 0.045$. Although these differences may seem significant, we point out that: (i) they would be smaller for any values of $\theta \in [0, \, \pi/2]$ and (ii) they are only reached in correspondence of the two periastron passages, while they remain much smaller over the rest of the orbit. Hence we are relatively confident that there will be no significant changes in the best-fit parameters we found for different initial inclinations of S2. In addition,  \citet{GRAVITY:2019tuf} showed that also the inclination of Sgr~A*'s spin with respect to the observer frame plays an important role in the effects the cloud has on S2 motion. 
Indeed, results including the motion of other S-stars and Sgr~A*'s spin direction are left for future works.

Recently, \citet{Sengo:2022jif} studied constraints on scalar structures using EHT data. Not surprisingly, bounds are of order $\Lambda \sim 10\% M$, compatible with the measurement precision of the telescope. Our results improve consistently and considerably this estimate for Sgr~A*, showing that a bosonic structure can only exist with a maximum (fractional) mass of $\Lambda \approx 10^{-3}$, at least for spin 0 fields.

\citet{Yuan:2022nmu} used the motion of S2 to derive an upper limit of $\delta M \lesssim 10^{-4} \, M$ for a scalar cloud with particle mass $m_s = 10^{-18} \, \rm eV$ ($\alpha \sim 0.015$) interacting with either the Higgs boson or the photon. Their estimate only uses publicly available and not GRAVITY data, which, due to their very small uncertainties, dominate our likelihood. This is reflected in the best-fit parameters found which are not compatible (within $3 \sigma$ uncertainties) with the most recent ones reported in \citet{GRAVITY:2021xju}. We argue that this difference already at the non-perturbative level may lead to misleading results when the cloud is included in the fit.

Finally, we point out that the spin of Sgr~A* is relevant when discussing superradiant phenomena, since it affects the possible origin of the scalar cloud. Despite a recent work by \citet{Fragione:2020khu} placing a strong constraint on Sgr~A* spin parameter ($\chi \lesssim 0.1$), other studies \citep{Qi:2020xzi} question such result, and show that the current astrometric measurements are yet not sufficient to constrain the value of the spin. On the other hand, \citet{Kato:2009zw} used quasi-periodic oscillations in the radio emissions of Sgr~A* to claim that its spin is $\chi = 0.44 \pm 0.08$. The current best estimate for Sgr~A*'s spin comes from the EHT observations \citep{Broderick:2010kx}, which reported a measurement of $\chi = 0.00 \pm 0.64$ where the error is the $1 \sigma$ uncertainty. 
 Due to the high uncertainty of these results and the ongoing discussion about it, it can be assumed without loss of generality that Sgr~A* is (was) in fact spinning enough to engage a superradiant instability. We note, however, that even a non-spinning BH can bind a scalar ``cloud'' if it was grown via some other mechanism (for example, primordial, ~\citet{Cardoso:2022vpj}).

An upgrade of the Gravity experiment towards Gravity+ is ongoing at the time of writing, as well as the commissioning of the ERIS instrument. The increased sensitivity of Gravity+ and the patrol field of view of ERIS strongly increase the prospects of detecting and tracking further stars in inner orbits, putting stronger constraints on the scalar cloud.

\section*{Acknowledgements}
We are very grateful to our funding agencies (MPG, ERC, CNRS [PNCG, PNGRAM], DFG, BMBF, Paris Observatory [CS, PhyFOG], Observatoire des Sciences de l'Univers de Grenoble, and the Funda\c c\~ao para a Ci\^encia e a Tecnologia), to ESO and the Paranal staff, and to the many scientific and technical staff members in our institutions, who helped to make NACO, SINFONI, and GRAVITY a reality. 
V.C.\ is a Villum Investigator and a DNRF Chair, supported by VILLUM Foundation (grant no. VIL37766) and the DNRF Chair program (grant no. DNRF162) by the Danish National Research Foundation.
V.C.\ acknowledges the financial support provided under the European
Union's H2020 ERC Advanced Grant ``Black holes: gravitational engines of discovery'' grant agreement
no.\ Gravitas–101052587. Views and opinions expressed are, however, those of the author only and do not necessarily reflect those of the European Union or the European Research Council. Neither the European Union nor the granting authority can be held responsible for them.
This project has received funding from the European Union's Horizon 2020 research and innovation programme under the Marie Sklodowska-Curie grant agreement No 101007855.
We acknowledge the financial support provided by FCT/Portugal through grants 
2022.01324.PTDC, PTDC/FIS-AST/7002/2020, UIDB/00099/2020 and UIDB/04459/2020.

\section*{Data Availability}
Publicly available data for astrometry and radial velocity up to 2016.38 can be found in Table 5 the electronic version of \cite{2017ApJ...837...30G} at this link:  \url{https://iopscience.iop.org/article/10.3847/1538-4357/aa5c41/meta#apjaa5c41t5}.



\bibliographystyle{mnras}
\bibliography{biblio} 




\appendix

\section{Details about numerical integration}
\label{app:num_integration}

The numerical integration of the equation of motion in \eqref{true_eom} is performed making use of the Python library \textbf{scipy.integrate.solve\_ivp} with a Runge-Kutta 5(4) algorithm, meaning that the steps are evaluated using a 5-th order method while the error is controlled assuming the accuracy of the 4-th order method. The convergence of the integration is assured by looking at the conservation of energy over the entire integration period (almost two orbits in $\sim 30$ years gives $\Delta E/E \sim \mathcal{O}(10^{-10})$).

Kepler's equation is solved instead using a Python's root finder (\textbf{scipy.optimize.newton}) which implements a Newton-Raphson method. The latter solves the equation with precision of $\mathcal{O}(10^{-16})$. 
\section{Coordinates transformations and inclusion of relativistic effects.}
\label{app:relativistic_effects}

\subsection{Coordinate transformation}
\label{app:coord_transf}

The transformation from the orbital reference frame to the observer reference frame can be achieved using the following conversion:
\beq
\begin{split}
& x' = A x_{\rm BH} + F y_{\rm BH} \,\,\,\,\,\,\,\,\,\,\,\,\,\,\,\,\,\,\,\,\, v_{x'} = A v_{x_{\rm BH}} + F v_{y_{\rm BH}} \\
& y' = B x_{\rm BH} + G y_{\rm BH}\,\,\,\,\,\,\,\,\,\,\,\,\,\,\,\,\,\,\,\,\, v_{y'} = B v_{x_{\rm BH}} + G v_{y_{\rm BH}} \\
& z_{\rm obs} = -(C x_{\rm BH} + H y_{\rm BH}) \,\,\,\,\,\,\,\,\,\,\,\, v_{z_{\rm obs}} = -(C v_{x_{\rm BH}} + H v_{y_{\rm BH}}) \, ,
\end{split}
\label{coord_obs}
\eeq 
where $A, B, C, F, G, H$ are the Thiele-Innes parameters \citep{Catanzarite:2010wa} defined as:
\begin{equation}
\begin{split}
& A = \cos \Omega \cos \omega -\sin \Omega \sin \omega \cos i \\
& B = \sin \Omega \cos \omega + \cos \Omega \sin \omega \cos i \\
& F = -\cos \Omega \sin \omega - \sin \Omega \cos \omega \cos i \\
& G = -\sin \Omega \sin \omega + \cos \Omega \cos \omega \cos i \\
& C = - \sin \omega \sin i \\
& H = -\cos \omega \sin i \,, 
\end{split}
\end{equation}
while the Cartesian coordinates $\{x_{\rm BH}, y_{\rm BH}, z_{\rm BH}\}$ and velocities $\{v_{x_{\rm BH}}, v_{y_{\rm BH}}, v_{z_{\rm BH}}\}$ are those obtained from the numerical integration. For a more detailed discussion about how the coordinate system $\{x', y', z_{\rm obs}\}$ and the above transformation are defined we refer the reader to Figure 1 and Appendix B of \citet{Grould:2017bsw}. 

\subsection{Relativistic effects and R{\o}mer's delay}

As said in the main text, there are two main contributions that must be taken in consideration when S2 approaches the periastron: the relativistic Doppler shift and the gravitational redshift. Both of them induce a shift in the spectral lines of S2 that affects the radial velocity measurments. 
The former is given by
\beq 
1 + z_{D} = \frac{1 + v_{z_{\rm obs}}}{\sqrt{1- v^2}} \,,
\eeq 
while the gravitational redshift is defined as 
%
\beq 
1 + z_{\rm G} = \frac{1}{\sqrt{1 - 2 M/r_{\rm em}}} \,.
\eeq
The two shifts can be combined using Eq.(D.13) of \citet{Grould:2017bsw} to obtain the total radial velocity 
\beq
V_R \approx \frac{1}{\sqrt{1 - \epsilon}} \cdot \frac{1 + v_{z_{\rm obs}}/\sqrt{1-\epsilon}}{\sqrt{1 - v^2/(1 - \epsilon)}} - 1 \,.
\eeq
where $\epsilon = 2M/r_{\rm em}$. In the total space velocity $v = |\textbf{v}|$ we must also add a correction due to the Solar System motion. We followed the most recent work of \citet{2020ApJ...892...39R} and take a proper motion of Sgr~A* of 
\beq
\begin{split}
& v_x^{\rm SSM} = -5.585 \, \rm mas/yr = 6.415 \cos(209.47^{\circ}) \, mas/yr \, ,\\
& v_y^{\rm SSM} = -3.156 \, \rm mas/yr = 6.415 \sin(209.47^{\circ}) \, mas/yr \, .
\end{split}
\eeq
The R{\o}mer's delay is instead included using the first order Taylor's expansion of Eq.~\eqref{roemer_equation}, which reads:
\begin{equation}
    t_{\rm em} = t_{\rm obs} - \frac{z_{\rm obs}(t_{\rm obs})}{1 + v_{z_{\rm obs}}(t_{\rm obs})} \,.
    \label{t_em}
\end{equation}
\noindent The difference between the exact solution of Eq.~\eqref{roemer_equation} and the approximated one in \eqref{t_em} is at most $\sim 4$ s over S2 orbit and therefore negligible. The R{\o}mer effect affects both the astrometry and the spectroscopy, with an impact of $\approx 450 \, \mu$as on the position and $\approx 50$ km/s at periastron for the radial velocity. Our results recover the previous estimates for this effect in \citet{Grould:2017bsw, GRAVITY:2018ofz}.

\section{Corner plots}

Here we report the corner plots for two representative values of $\alpha$ ($\alpha = 0.01$ and $\alpha = 0.001$), to show the behaviour of the parameters when the cloud is located in and outside S$2$'s orbital range. The strong correlation between $\Lambda$ and the periastron passage $t_p$ when $\alpha = 0.01$ can be understood following the argument of \cite{Heissel:2021pcw}: the presence of an extended mass will induce a retrograde precession in the orbit that will result in a positive shift of the periastron passage time, needed to compensate the (negative) shift in the initial true anomaly. Indeed, when considering the Schwarzschild precession, which instead induces a prograde precession (hence a positive initial shift in the true anomaly), $t_p$ will undergo a negative shift, as can be seen from the strong anti-correlation between $f_{\rm SP}$ and $t_p$ reported in \cite{GRAVITY:2020gka}.

\begin{figure*}
\includegraphics[width=\textwidth]{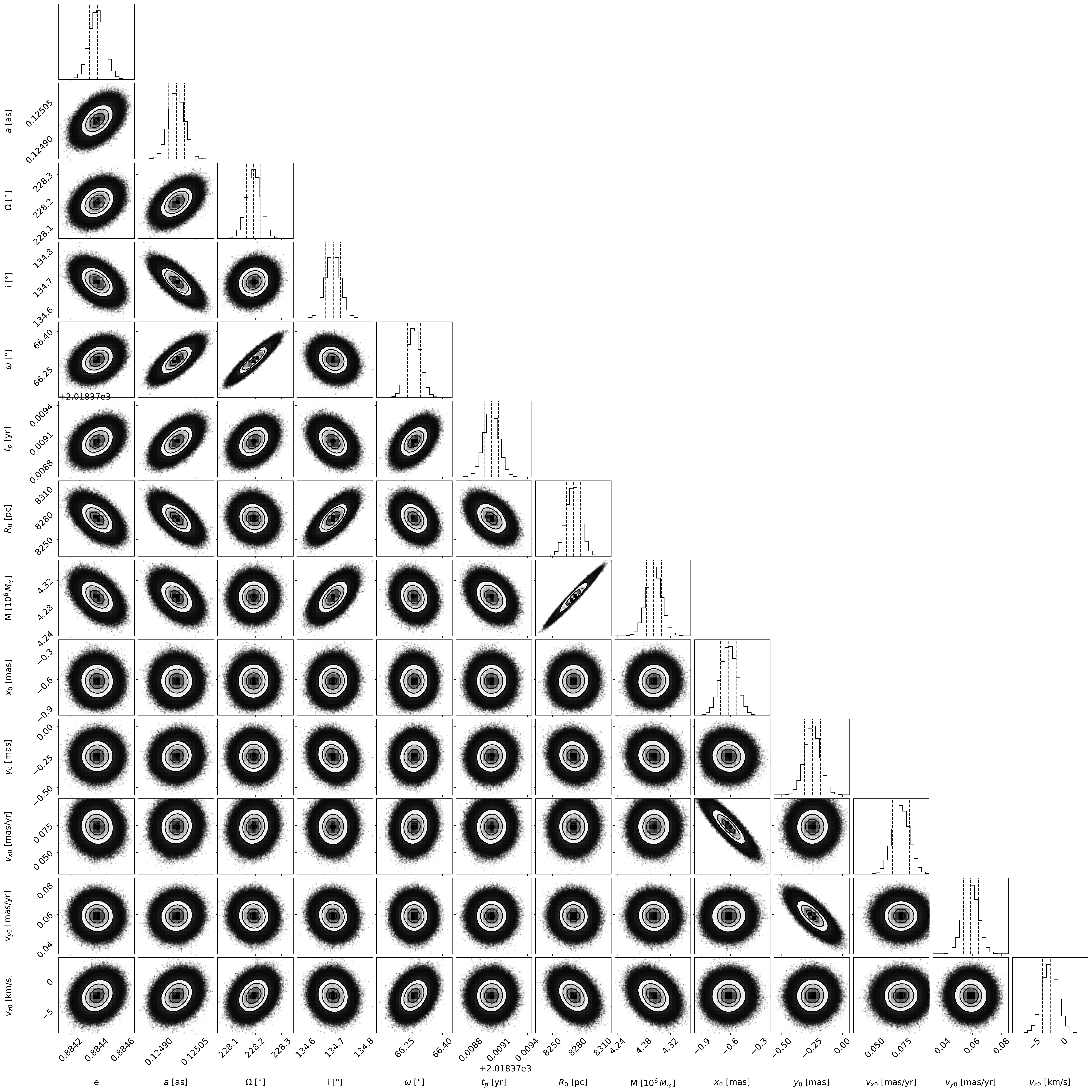}
    \caption{Corner plot of the fitted parameters with $f_{\rm SP} = 1$ and $\alpha = 0.01$. Dashed lines represent the $0.16$, $0.50$ and $0.84$ quantiles of the distributions.}
    \label{fig:corner_plot_alpha1}
\end{figure*}
\newpage
\begin{figure*}
\includegraphics[width=\textwidth]{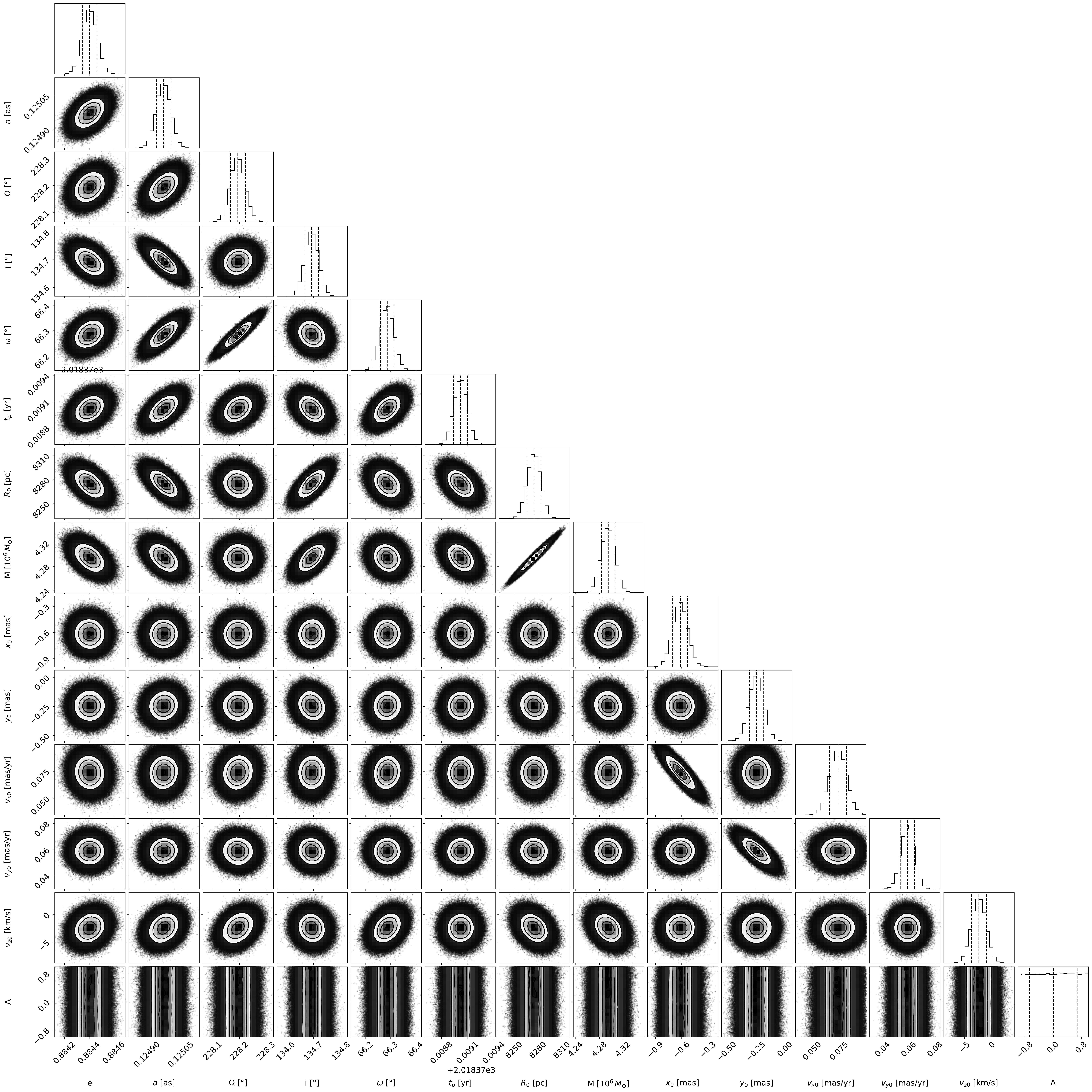}
    \caption{Corner plot of the fitted parameters with $f_{\rm SP} = 1$ and $\alpha = 0.001$. Dashed lines represent the $0.16$, $0.50$ and $0.84$ quantiles of the distributions.}
    \label{fig:corner_plot_alpha2}
\end{figure*}

\bsp	
\label{lastpage}
\end{document}